\documentclass[aps,reprint]{revtex4-2}
\usepackage{etoolbox} % for \appto
\usepackage{blindtext}
\usepackage{graphicx}
\usepackage{subcaption}
\usepackage{multirow}%
\usepackage{amssymb}
\usepackage{xcolor}
\usepackage{graphicx}
\usepackage{floatrow}
\usepackage[noend]{algpseudocode}
\usepackage[fleqn]{amsmath}
\usepackage[font=small,skip=3pt]{caption}
\usepackage{makecell}
\usepackage{textcomp}%
\usepackage{hyperref} 
\usepackage{cleveref}
\usepackage{url}
\usepackage{tikz}
\usepackage{siunitx}
\DeclareSIUnit\MWh{MWh}
\DeclareSIUnit\dollar{\mathdollar}
\usepackage[version=4]{mhchem} % added by JAS for typesetting isotopes
\usepackage{booktabs} % added by JAS for tables
\usepackage{bm}
\let\mrm\mathrm
\makeatletter
\appto{\appendix}{%
  \@ifstar{\def\theequation@prefix{A.}}%
          {}%!TEX encoding = UTF-8 UnicodeX
}
\makeatother

\newcolumntype{B}{>{\color{red}}c}

\begin{document}
\title{Isotope Production in Muon-Catalyzed-Fusion Systems}
\author{J. F. Parisi$^{1}$}
\email{jason@marathonfusion.com}
\author{A. Rutkowski$^{1}$}
\affiliation{$^1$Marathon Fusion, 150 Mississippi, San Francisco, 94107, CA, USA}

\begin{abstract}
Producing valuable isotopes with high-flux high-energy neutrons generated by muon-catalyzed fusion ($\mu$CF) reactions could substantially improve the economic prospects for muon-catalyzed fusion. Because no external heating is required for $\mu$CF, heat flux constraints are significantly relaxed compared with fusion systems requiring external heating. This could allow $\mu$CF to attain much higher neutron flux without breaching material heat flux limits. If muon production rates can be increased, $\mu$CF systems employing transmutation could be viable well before energy breakeven is possible. For $\mu$CF systems transmuting valuable isotopes, the required number of catalyzed fusion events per muon and muon energy generation cost can be relaxed by several orders of magnitude relative to electricity-generating systems, making $\mu$CF an attractive high-flux neutron source. We show an example $\mu$CF system with a 10 gram ${}^{226}\mathrm{Ra}$ feedstock and a steady-state muon rate of $10^{12}$ muons / second - roughly half a kilowatt of fusion power - could produce 20 mg of ${}^{225}\mathrm{Ac}$ per year - comparable to 400 times global supply in 2024. As higher muon rate sources become available, many other radioisotope transmutation pathways become viable. These findings motivate the accelerated development of $\mu$CF systems for neutron-driven isotope production far before net energy generation is possible.
\end{abstract}

\maketitle

\section{Introduction}

In this work, we show how muon-catalyzed fusion ($\mu$CF) can serve as an intense neutron source for making valuable radioisotopes. With isotope production, the requirements are significantly reduced for the production energy and number of catalyzed fusion reactions per muon compared to power generating $\mu$CF systems.

In a D-T mixture, formation of the muonic $(d\mu t)^+$ molecular ion enables rapid fusion to a 3.5 MeV $\alpha$ and a 14.1 MeV neutron \cite{frank1947hypothetical,Alvarez1957,Jones1989,Ponomarev1990,Froelich1992} with a total cycle time $t_\mrm{c}$. The muon is usually released to catalyze additional fusions, but two mechanisms limit repeated cycling: the $\tau =$ 2.2 $\mu$s muon lifetime and the $p\sim 0.5\%$ $\alpha$-sticking probability that binds the muon to the product nucleus \cite{Jackson1957,Petitjean1989}. Experiments have achieved $t_\mrm{c}$ values corresponding to $N_{\mathrm{fus,\mu}} =  1 /(t_\mrm{c}/\tau + p) \simeq 150$ catalyzed reactions per muon \cite{Jones1989} corresponding to  $t_\mrm{c} \sim 4$ ns. Each D-T cycle combined with neutron multiplication and $\ce{^6Li}$(n,$\alpha$)t (+4.8 MeV) releases at most $E_\mrm{comb} \approx 20.4$ MeV \cite{sawan_physics_2006} per cycle. Therefore, each muon can generate at most $E_{\mathrm{fus,\mu}} = N_{\mathrm{fus,\mu}} E_\mrm{comb} \approx 3.1$ GeV of energy. 

The 3.1 GeV energy yield per muon is close to, but still below, the level required for energy breakeven. Current muon sources can produce a single muon via pion decay with an effective cost of $E_\mu \sim 5$ GeV per muon \cite{Jandel1989}, although there are proposals for reducing $E_\mu$ \cite{iiyoshi2019muon,Kelly2021} and increasing $N_\mrm{fus,u}$ \cite{froelich1994fusion,kino1996muonic,iiyoshi2019muon}. Recent modeling \cite{Newburg2025} suggests that active-target muon sources could achieve substantially lower energy cost, simulating 3.0 GeV per negative muon, but further technology development is required to demonstrate this benefit experimentally. Thus, in present systems the fusion energy released per muon is at best comparable to, and typically smaller than, the accelerator energy invested per muon. Once realistic wall-plug efficiencies and auxiliary loads are included, $\mu$CF experiments remain below energy breakeven even under favorable conditions \cite{Jackson1957,Kelly2021}.

Although $\mu$CF does not yet support net-energy production, its neutron output is already nearly high enough to be useful for other applications such as isotope production. Each catalyzed D–T reaction yields a 14.1 MeV neutron without external plasma heating (and corresponding high wall heat flux loads). Recent work proposed using a $\mu$CF to transmute long-lived fission products \cite{iiyoshi2019muon,yamamoto2021transmutation}. In this work, we propose making valuable isotopes via fast neutron-driven transmutation \cite{li2023feasibility,evitts2025theoretical,ponsard2014production,lederer2024measurement,parisi2025production,rutkowski2025scalable}. Conventional neutron sources such as fission reactors and spallation systems face constraints in achievable neutron flux and fluence, thermal loading, waste generation, and proliferation risks. D–T $\mu$CF offers an alternative accelerator-driven pathway to produce fusion neutrons at low temperature and at relatively favorable neutron to heat flux ratio.

%Recent advances in negative muon source intensity further motivate $\mu$CF as a neutron-driven isotope production platform. 

Producing high rates of negatively charged muons is challenging, with state-of-the-art facilities producing several $\sim 10^7$ $\mu^-$/s: J-PARC MUSE produces $\sim 2-4 \cdot 10^{7}$ $\mu^-$/s in a 25 Hz pulsed beam \cite{Kawamura2025_JPARC_MUSE}, PSI achieves $\sim 2 \cdot 10^{7}$ $\mu^-$/s continuous wave \cite{Chen2023_muE1_simulation}, and MuSIC reported $\sim 4 \cdot 10^{7}$ $\mu^-$/s in a direct current beamline \cite{cook2017delivering}. Upgrades at PSI \cite{dal2023future,Papa2023_PSIBeamDevelopments_Muon4Future,valetov2024beamline} and J-PARC \cite{Harada2024_JPARC_TS2_neutronic} are likely to push negative muon production rates above $10^8$ $\mu^-$/s. Nonetheless, despite these advances, higher muon beam rates are needed even for valuable isotope production. 

%\textcolor{red}{statement on Cu64} (omit for now)

%Modern accelerator systems routinely deliver $\sim 10^8$ muons/s~\cite{berger2014mu3e,cook2017delivering,delahaye2019muon}, with ongoing and projected upgrades targeting one to two orders of magnitude higher flux~\cite{prokscha2004new,delahaye2019muon,boscolo2019future,apyan2023gamma,cai2024towards}. While these intensities are far below the $10^{16}$–$10^{19}$ muons/s needed for megawatt to gigawatt scale fusion power, they are sufficient to drive high-value neutron-induced transmutations, including production of medical isotopes such as $\ce{^{225}Ac}$. In this regime, economic viability is determined by the value per neutron rather than by energy breakeven, so $\mu$CF can be attractive even when $N_{\mathrm{fus,\mu}}$ and $E_\mu$ remain below the requirements for net power production.

This work is structured as follows. In \Cref{sec:muon_cat_fus} we calculate the heat and neutron fluxes $\mu$CF systems. The transmutation breakeven condition is discussed in \Cref{sec:breakeven}. We describe a small system for $\ce{^225Ac}$ production in \Cref{sec:ac225}. We conclude in \Cref{sec:discussion}.

\section{Heat and Neutron Fluxes in Muon-Catalyzed Fusion} \label{sec:muon_cat_fus}

%\subsection{Fusion and Wall Power}

In $\mu$CF, a negative muon binds to a hydrogen isotope, forming a muonic atom with a Bohr radius $\sim 200$ times smaller than that of an electron. The muon brings two nuclei into close proximity, enabling nuclear tunneling and fusion at low temperatures.

\begin{figure*}[tb!]
    \centering
    \begin{subfigure}[t]{0.49\textwidth}
    \centering
    \includegraphics[width=1.0\textwidth]{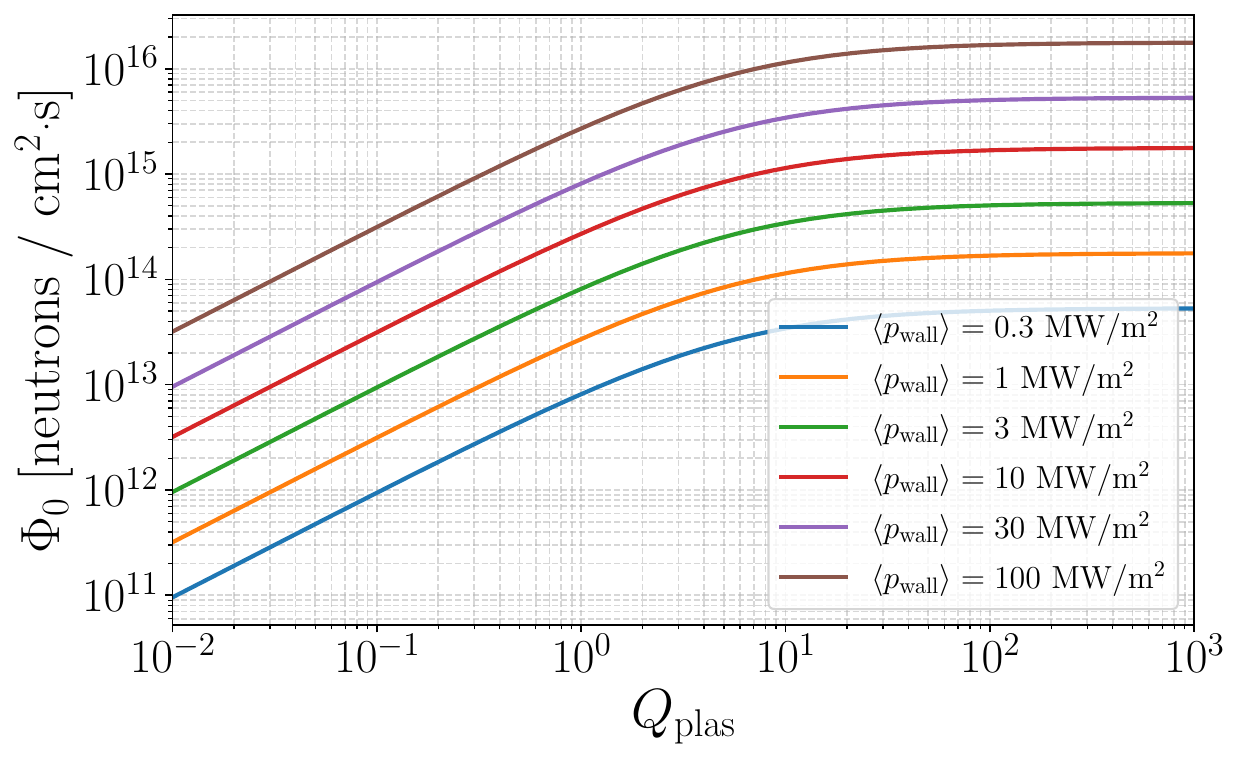}
    \caption{}
    \end{subfigure}
    \centering
    \begin{subfigure}[t]{0.49\textwidth}
    \centering
    \includegraphics[width=1.0\textwidth]{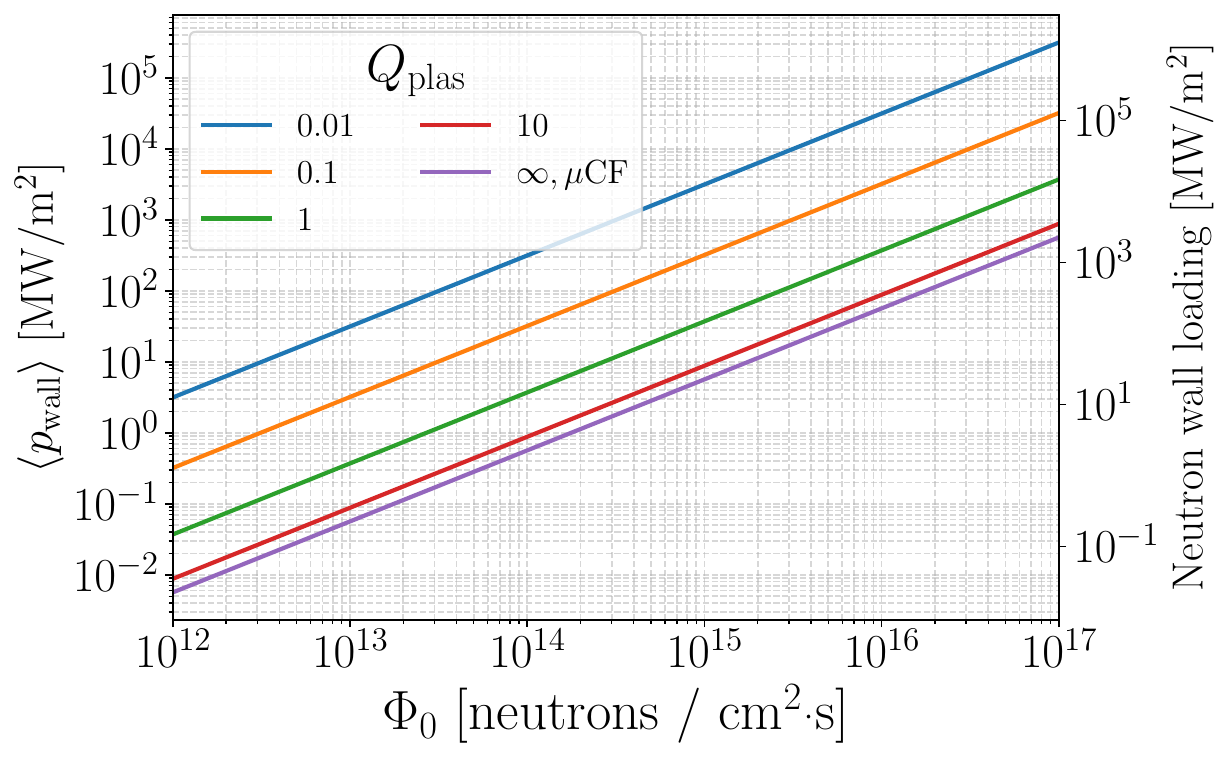}
    \caption{}
    \end{subfigure}
    \caption{Heat flux and neutron wall loading versus neutron flux for a range of plasma gains, including a $\mu$CF system. We use $\eta_\mrm{abs} = 0.90$.}
    \label{fig:pwall_versus_flux}
\end{figure*}

The muon-catalyzed D-T reaction is
\begin{equation}
\mu^- + \mathrm{D} + \mathrm{T} \;\longrightarrow\; (\mathrm{D}\mathrm{T}\mu)^+ \;\longrightarrow\;
{}^{4}\mathrm{He} + n + \mu^- + 17.6 \mathrm{MeV}.
\label{eq:muon_DT_cycle}
\end{equation}
Each muon on average catalyzes $N_\mrm{fus,\mu}$ fusion reactions before decay or sticking to a helium nucleus. Each muon has an energy production cost $E_\mu$. Modern muon sources can achieve $E_\mu \simeq 5$ GeV \cite{Jones1989}. The total fusion energy released from all fusions catalyzed by a single muon is therefore
\begin{equation}
    E_\mrm{fus,\mu} = N_\mrm{fus,\mu} E_\mrm{fus},
    \label{eq:Emu_total}
\end{equation}
where $E_\mrm{fus}=17.6$ MeV. The fusion rate in a D-T target is
\begin{equation}
    \dot{N}_\mrm{n} = R_\mu N_\mrm{fus,\mu},
    \label{eq:Ndotn}
\end{equation}
where $R_\mu$ is the number of muons absorbed per second. Details on deriving \Cref{eq:Ndotn} are provided in Appendix \ref{app:muoncf}. The total power to the wall of a muon-catalyzed fusion machine is
\begin{equation}
    P_\mrm{wall} = P_\mrm{\alpha},
    \label{eq:Pwall_muCF}
\end{equation}
where $P_\alpha$ is the alpha particle heating power. For this work, we neglect the energy deposition of muons in the target, which can be substantial ($\sim$MeV) in some cases \cite{kammel2021musun}. \Cref{eq:Pwall_muCF} is qualitatively different to fusion machines that heat D-T fuel with external sources. The average power per unit area to the wall in $\mu$CF is given by the fusion alpha power,
\begin{equation}
    \langle p_\mrm{wall} \rangle^\mu \equiv \frac{P_\mrm{wall}}{A_b} = E_\mrm{fus} \Phi_0 f_\alpha,
    \label{eq:pwall_initial}
\end{equation}
where $A_b$ is the area of the inner wall of the blanket surrounding the fuel and $\Phi_0$ is the average first-wall neutron flux,
\begin{equation}
    \Phi_0 \equiv \frac{\dot{N}_\mathrm{n}}{A_b},
\end{equation}
and $f_\alpha = 1/5$ is the fusion power fraction carried by alpha particles. Crucially, the wall loading in a $\mu$CF system is equivalent to an `ignited' heated plasma experiment with $Q_\mrm{plas} = \infty$,
\begin{equation}
    \langle p_\mrm{wall} \rangle \equiv \frac{P_\mrm{wall}}{A_b} = E_\mrm{fus} \Phi_0 \left(  f_\alpha  + \frac{1}{ \eta_\mrm{abs} Q_\mrm{plas}} \right),
    \label{eq:pwall_initial2}
\end{equation}
where $\eta_\mrm{abs}$ is the plasma heating absorption efficiency,
\begin{equation}
\eta_\mrm{abs} \equiv \frac{P_\mrm{abs}}{P_\mrm{heat}},
\label{eq:etaabs}
\end{equation}
$P_\mrm{abs}$ is the total absorbed heating power in a plasma, $P_\mrm{heat}$ is the heating power, and $Q_\mrm{plas}$ is the plasma gain \cite{Lawson1957,Wurzel2022}
\begin{equation}
    Q_\mrm{plas} \equiv \frac{P_\mrm{fus}}{ P_\mrm{abs}} = \frac{P_\mrm{fus}}{\eta_\mrm{abs} P_\mrm{heat}}.
    \label{eq:Qplas}
\end{equation}
The total fusion power in a $\mu$CF system is
\begin{equation}
    P_\mathrm{fus} = R_\mu N_\mrm{fus,\mu} E_\mathrm{fus},
\end{equation}
with the corresponding muon production power requirement
\begin{equation}
    P_\mu = R_\mu E_\mu / f_\mrm{stop},
    \label{eq:P_mu_eq}
\end{equation}
where $f_\mrm{stop}$ is the fraction of muons that stop in the target. Rearranging \Cref{eq:pwall_initial2} for $\Phi_0$ gives
\begin{equation}
    \Phi_0 = \frac{\langle p_\mrm{wall} \rangle}{E_\mathrm{fus}} \frac{1}{f_\alpha + \frac{1}{\eta_\mathrm{abs} Q_\mathrm{plas}}}.
\end{equation}
In \Cref{fig:pwall_versus_flux} we plot $\langle p_\mrm{wall} \rangle$ in versus $Q_\mathrm{plas}$ and $\Phi_0$: while transmutation might be economically attractive in externally heated systems at low $Q_\mathrm{plas}$, a big challenge is that $\langle p_\mrm{wall} \rangle$ can be prohibitively large at high $\Phi_0$ \cite{parisi2025isotope}. In $\mu$CF, $\langle p_\mrm{wall} \rangle^\mu$ is always smaller than the equivalent in a heated fusion system $\langle p_\mrm{wall} \rangle$, except for when $Q_\mathrm{plas}=\infty$ in a heated system, giving $\langle p_\mrm{wall} \rangle^\mu = \langle p_\mrm{wall} \rangle$.

\section{Transmutation Breakeven Condition} \label{sec:breakeven}

In this section we calculate the breakeven condition for a $\mu$CF system producing isotopes.

\begin{figure*}[tb!]
    \centering
    \begin{subfigure}[t]{0.49\textwidth}
    \centering    
    \includegraphics[width=0.99\textwidth]{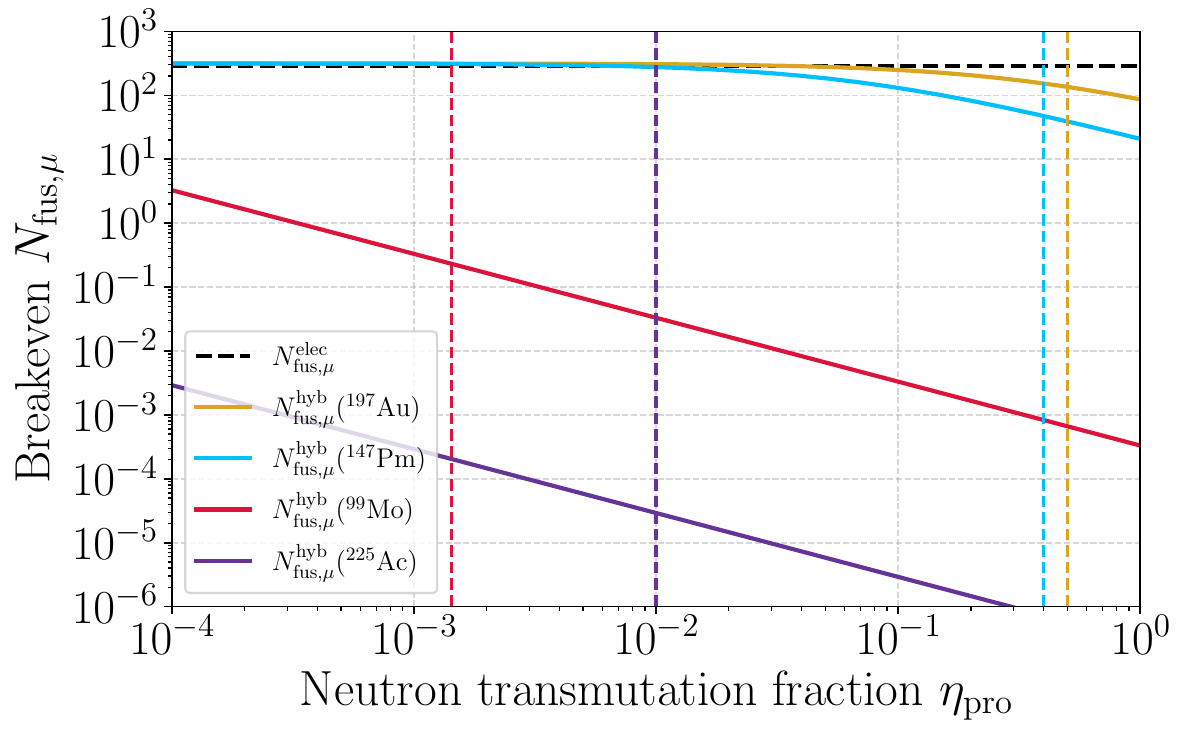}
    \caption{}
    \end{subfigure}
    \centering
    \begin{subfigure}[t]{0.49\textwidth}
    \centering    
    \includegraphics[width=0.99\textwidth]{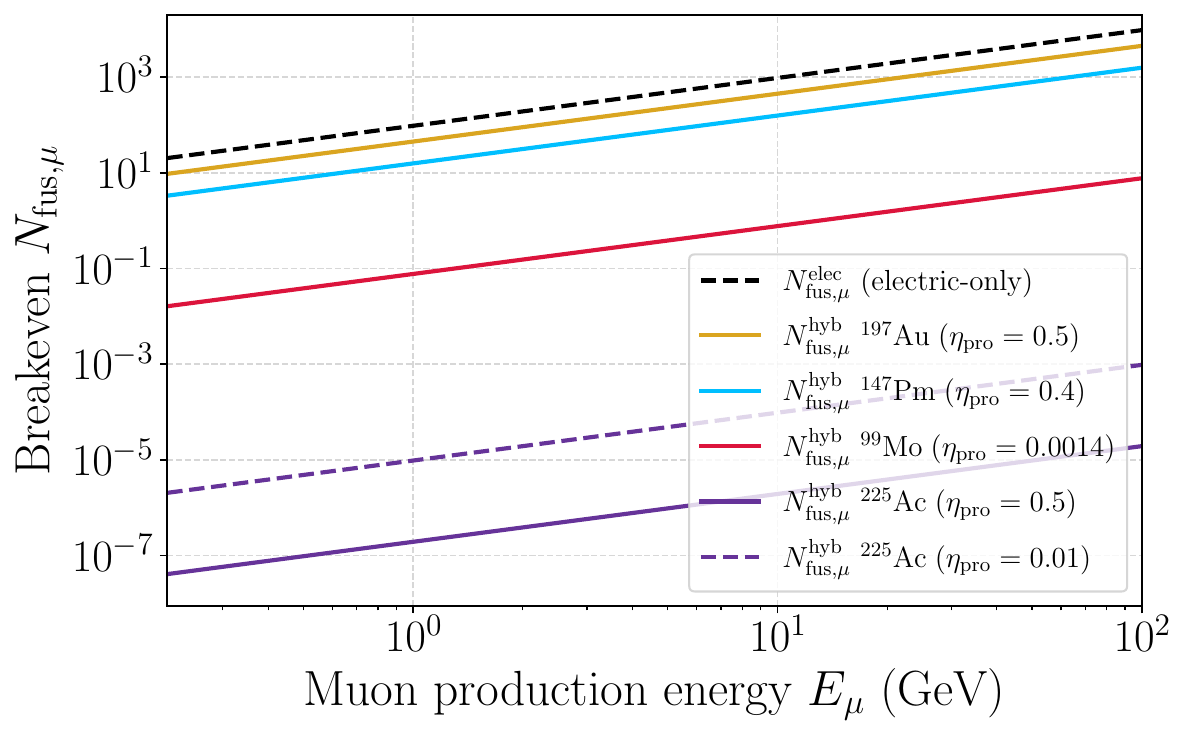}
    \caption{}
    \end{subfigure}
    \caption{(a) Required number of catalyzed fusion reactions per muon $N_\mrm{fus,\mu}$ assuming $E_\mu$ = 3 GeV / muon (see \Cref{eq:N_mu_breakeven_elec,eq:N_mu_breakeven_transmuter}) versus neutron transmutation fraction $\eta_\mrm{pro}$. Vertical dashed lines are typical capture fractions for each isotope in a fusion transmutation blanket. (b) Breakeven $E_\mrm{\mu}$ versus $\eta_\mrm{pro}$ for each isotope. We use $\eta = 0.4$, $\eta_\mu = 0.8$, $\kappa = 1.1$, $\xi_\mrm{circ} = 0.1$, $\xi_\mrm{pro}$ = 0.1, $C_e = \$50$/MWh. We assumed $C_\mrm{pro} = 1.4 \cdot 10^5$ \$/kg for $\ce{^197Au}$, $C_\mrm{pro} = 10^6$ \$/kg for $\ce{^147Pm}$, $C_\mrm{pro} = 5 \cdot 10^{11}$ \$/kg for $\ce{^99Mo}$, and $C_\mrm{pro} = 5 \cdot 10^{14}$ \$/kg for $\ce{^225Ac}$.}
    \label{fig:number_of_muonic_reactions}
\end{figure*}

The net revenue $\dot{R}$ (\$/s) from a $\mu$CF system that sells isotopes (and possibly electricity) is
\begin{equation}
    \dot{R} = \dot{M}_\mrm{pro} C_\mrm{pro} + \left( P_\mathrm{e} - P_\mathrm{pro} - P_\mu \right) \tilde{C}_\mrm{e}.
    \label{eq:governing_eq_f}
\end{equation}
Here, $\dot{M}_\mrm{pro}$ is the transmutation rate in kg/s, $C_\mrm{pro}$ is the sale price of transmuted product in \unit{\dollar\per\kilo\gram}, $P_\mrm{e}$ is the net electric power, $P_\mathrm{pro}$ is the electric power for all transmutation and extraction systems in watts, and $\tilde{C}_\mrm{e} = C_\mrm{e} / 10^6 T_\mrm{hour}$ is the sale price of electricity in \unit{\dollar\per\joule}, where $C_\mrm{e}$ is the sale price of electricity in \unit{\dollar\per\MWh} and $T_\mrm{hour}=\qty{3600}{s}$. The mass production rate of transmuted product is
\begin{equation}
    \dot{M}_\mathrm{pro} \equiv \dot{N}_\mathrm{pro} m_\mathrm{pro} = \eta_\mathrm{pro} P_\mathrm{fus} \frac{m_\mathrm{pro}}{E_\mrm{fus}}.
    \label{eq:Mdotpro}
\end{equation}
where $\dot{N}_\mathrm{pro}$ is the isotope production rate, $m_\mathrm{pro}$ is the mass of a single product atom and $\eta_\mathrm{pro} = \dot{N}_\mrm{pro}/ \dot{N}_\mrm{n}$. A simple way to model the net electric power output $P_\mathrm{e}$ is
\begin{equation}
    P_\mathrm{e} = \eta ( P_\mathrm{th} + \eta_\mu P_\mu) - P_\mathrm{circ} ,
    \label{eq:Pe_notritium}
\end{equation}
where $\eta$ is the electricity conversion efficiency, $\eta_\mu$ is the fraction of beam power $P_\mu$ generating heat that can be generated into electricity, and $P_\mathrm{circ}$ is the recirculating power for all non-transmutation and non-muon source systems.  We model the total thermal power $P_\mathrm{th}$ using $P_\mathrm{th} = \mathcal{K} P_\mathrm{fus}$ where $\mathcal{K} \equiv \mathcal{M} f_\mathrm{n} + f_\alpha$, and $\mathcal{M}$ is the power multiplication in the blanket; studies have found $\mathcal{M} \simeq 1 - 1.2$ \cite{sawan_physics_2006}. $\mathcal{M}$ describes the net effect of blanket heat sources and sinks, including effects as neutron slowing, exothermic ${}^6\mrm{Li}$ reactions, endothermic $(\mathrm{n},\mathrm{2n})$ reactions, gamma ray absorption, and neutron losses. The power fraction carried by neutrons is $f_\mathrm{n} = 4/5$.  The total revenue is
\begin{equation}
    \dot{R} = P_\mathrm{th} \tilde{C}_\mathrm{e} 
        \left( \eta + \frac{\eta_\mathrm{pro}}{\mathcal{K}} 
        \frac{m_\mathrm{pro}}{E_\mathrm{fus}} 
        \frac{C_\mathrm{pro}}{\tilde{C}_\mathrm{e}} 
        - F_\mathrm{circ} - F_\mathrm{pro} - F_\mathrm{\mu} (1 - \eta \eta_\mu) \right),
    \label{eq:R_muCF}
\end{equation}
where the recirculating power fraction for the non-transmutation and transmutation systems are
\begin{equation}
    F_\mrm{circ} \equiv \frac{P_\mrm{circ}}{P_\mathrm{th}}, \;\;\;\;\;  F_\mrm{pro} \equiv \frac{P_\mrm{pro}}{P_\mathrm{th}}, \;\;\;\;\; F_\mu \equiv \frac{P_\mu}{P_\mathrm{th}}.
\end{equation}
By co-producing isotopes, the `revenue breakeven' condition usually associated with electricity \cite{Wurzel2022} becomes less stringent.

We will calculate a hybrid breakeven quantity $Q_\mrm{eng}^\mrm{hyb}$ that considers the parasitic power $P_{\mrm{pro}}$ needed by transmutation systems in addition to the additional revenue from isotope sales. Converting product revenue in \Cref{eq:R_muCF} to an electric-equivalent power gives,
\begin{equation}
    \widetilde P_{\mrm{pro}} = P_\mrm{th} \frac{\eta_\mathrm{pro}}{\mathcal{K}} 
        \frac{m_\mathrm{pro}}{E_\mathrm{fus}} 
        \frac{C_\mathrm{pro}}{\tilde{C}_\mathrm{e}}.
\end{equation}
A modified net `power' $P_{\text{net}}$ that includes the power-equivalent $\widetilde P_{\mrm{pro}}$ and the transmutation electricity cost $P_\mrm{pro}$ is 
\begin{equation}
  P_{\text{net}}
  = P_e
    + \widetilde P_{\mrm{pro}} - P_\mrm{pro} - P_\mu.
\end{equation}
To obtain `hybrid engineering' gain, divide by the total recirculating electric power $P_\mrm{circ}+P_\mrm{pro} + P_\mu$,
\begin{equation}
  Q_{\text{eng}}^{\text{hyb}}
  \;\equiv\;
  \frac{P_\mrm{net}}
       {P_{\text{circ}}+P_{\mrm{pro}}+ P_\mu},
  \label{eq:QengTrans}
\end{equation}
which reduces to the standard engineering gain \cite{Wurzel2022} when $P_\mrm{pro} \to 0$, $\tilde{P}_\mrm{pro} \to 0$, and $P_\mu \to 0$.

\begin{figure*}[tb!]
    \centering
    \begin{subfigure}[t]{0.48\textwidth}
    \centering
    \includegraphics[width=1.0\textwidth]{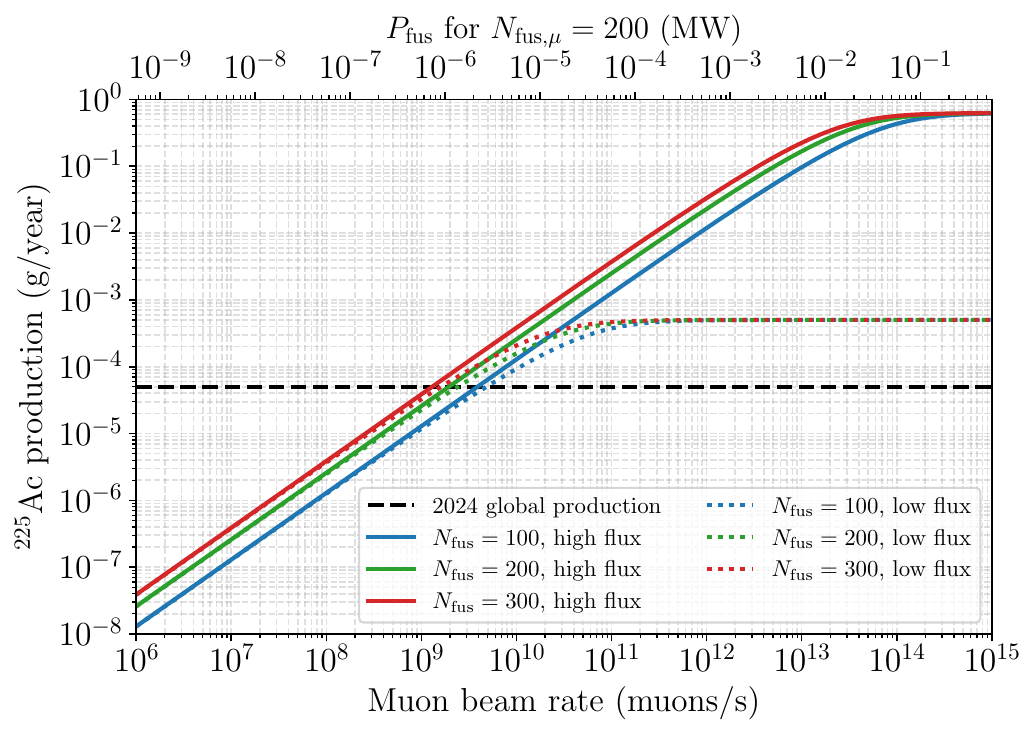}
    \caption{}
    \end{subfigure}
    \centering
    \begin{subfigure}[t]{0.48\textwidth}
    \centering
    \includegraphics[width=1.0\textwidth]{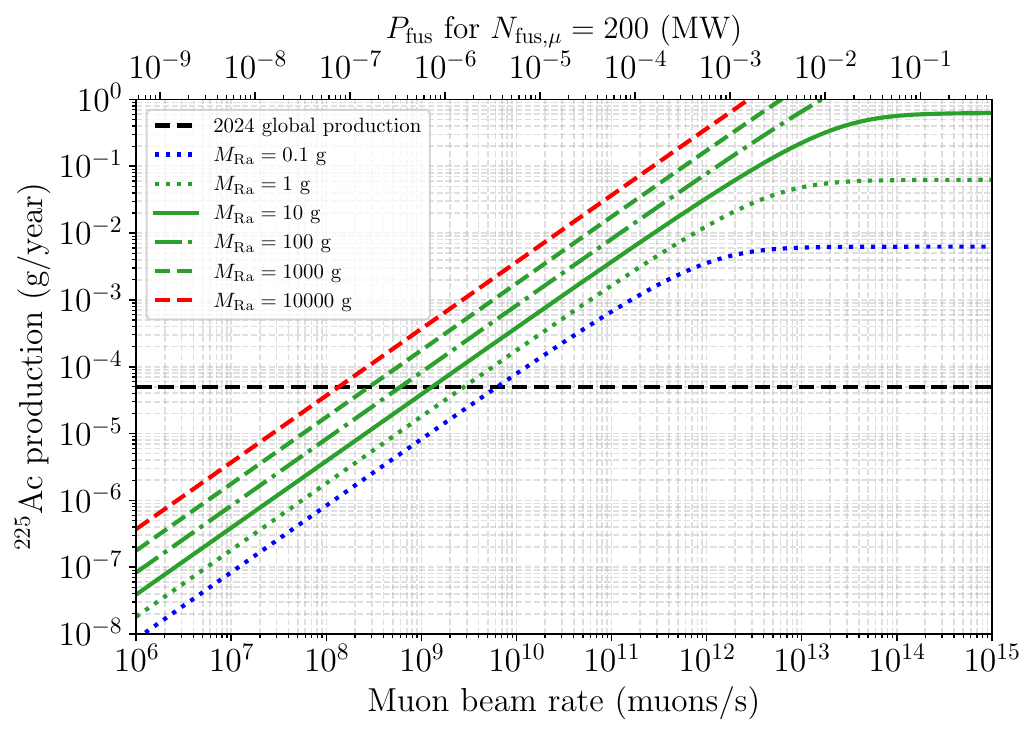}
    \caption{}
    \end{subfigure}
    \centering
    \begin{subfigure}[t]{0.48\textwidth}
    \centering
    \includegraphics[width=1.0\textwidth]{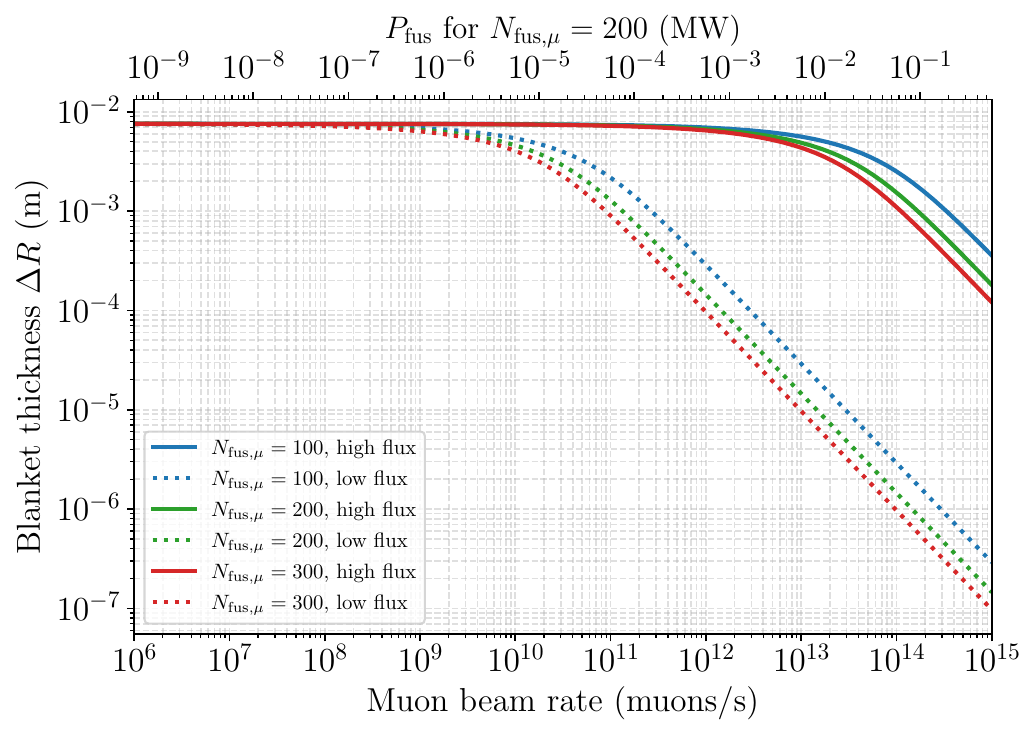}
    \caption{}
    \end{subfigure}
    \centering
    \begin{subfigure}[t]{0.48\textwidth}
    \centering
    \includegraphics[width=1.0\textwidth]{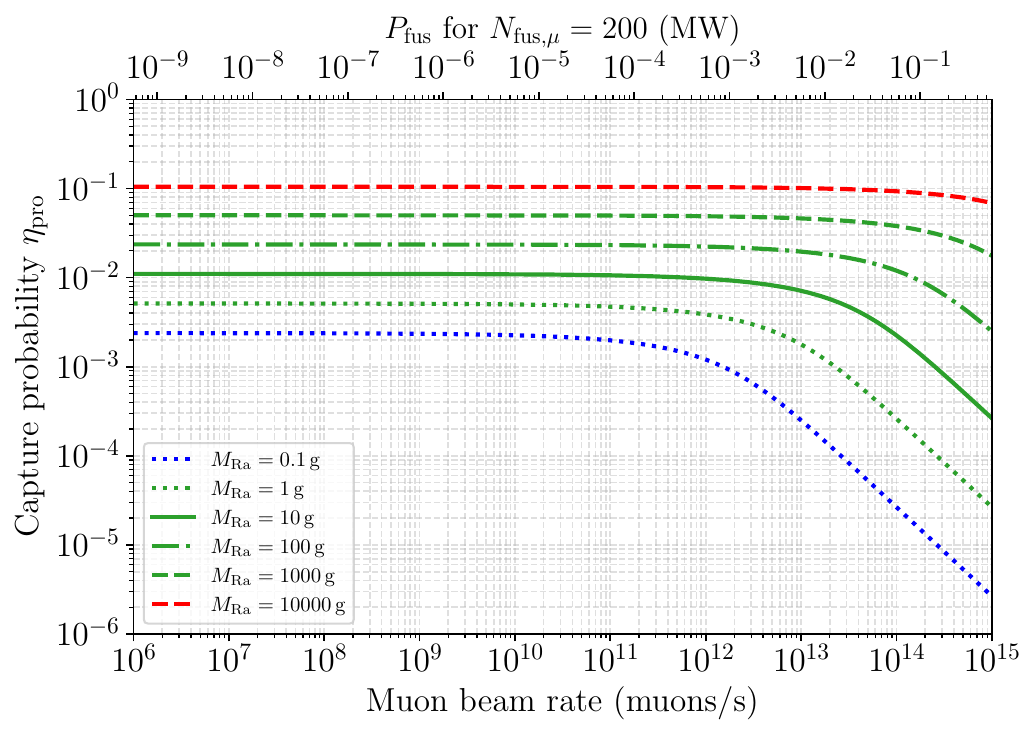}
    \caption{}
    \end{subfigure}
    \caption{$\ce{^225Ac}$ versus muon rate for a spherical $\mu$CF system with (a) High flux $2\cdot 10^{15}\ \mathrm{n\ cm^{-2}\ s^{-1}}$ and low flux $1.6\cdot 10^{12}\ \mathrm{n\ cm^{-2}\ s^{-1}}$ (with 10g $\ce{^226Ra}$ feedstock), (b) varying $\ce{^226Ra}$ feedstock mass (with $N_\mathrm{fus,\mu}$ = 200, high flux). (c) Blanket thickness $\Delta R$ (with 10g $\ce{^226Ra}$ feedstock), and (d) capture probability $\eta_\mathrm{pro}$ versus muon rate (with $N_\mathrm{fus,\mu}$ = 200, high flux). 2024 global production of $\ce{^225Ac}$ is reported at 3 Curies per year, or about 51 $\mu$g/yr \cite{Kraev2024PanTeraAc225}.}
    \label{fig:ac225_production_general}
\end{figure*}

We now calculate the modified plasma gain requirement by relating each parasitic term. We write
\begin{equation}
    P_\mrm{pro} = \xi_\mrm{pro} P_\mu,
\end{equation}
where we expect that $\xi_\mrm{pro} \ll 1$. Similarly,
\begin{equation}
     P_\mrm{circ} = \xi_\mrm{circ} P_\mu,
\end{equation}
where $\xi_\mrm{circ} \ll 1$. The hybrid engineering gain becomes
\begin{equation}
  Q_{\text{eng}}^{\text{hyb}}
  \;\equiv\;
  \frac{P_\mrm{net}}
       {P_\mu \left( 1 + \xi_\mrm{pro} + \xi_\mrm{circ} \right)}.
  \label{eq:QengTrans_v2}
\end{equation}
After some algebra, we find
\begin{equation}
    Q_{\text{eng}}^{\text{hyb}} = G - 1.
    \label{eq:Qenghyb_simpl}
\end{equation}
where
\begin{equation}
    G \equiv N_\mrm{fus,\mu} \frac{E_\mrm{fus}}{E_\mu}\frac{\left(1 - \eta \eta_\mu\right)^{-1}}{ 1 + \xi_\mrm{pro} + \xi_\mrm{circ}} \left( \eta_\mrm{pro} \frac{m_\mrm{pro}}{E_\mrm{fus}} \frac{C_\mrm{pro}}{\widetilde C_e} + \eta \mathcal{K} \right).
\end{equation}
%\begin{equation}
%    Q^*_\mrm{plas} = 1 / G.
%    \label{eq:Qstar_breakeven}
%\end{equation}
For a pure electric $\mu$CF FPP, $\xi_\mrm{pro} = C_\mrm{pro} = 0$, hybrid breakeven ($Q_{\text{eng}}^{\text{hyb}} \geq 0$) is
\begin{equation}
    N_\mrm{fus,\mu}^\mrm{elec} \geq  \frac{E_\mu}{E_\mrm{fus}} \frac{1}{\eta \mathcal{K}} \left( 1 + \xi_\mrm{circ} \right) \left(1 - \eta \eta_\mu\right).
    \label{eq:N_mu_breakeven_elec}
\end{equation}
With isotope production, hybrid breakeven is
\begin{equation}
     N_\mrm{fus,\mu}^\mrm{hyb} \geq  \frac{E_\mu}{E_\mrm{fus}} \frac{ 1 + \xi_\mrm{pro} + \xi_\mrm{circ}}{\eta_\mrm{pro} \frac{m_\mrm{pro}}{E_\mrm{fus}} \frac{C_\mrm{pro}}{\widetilde C_e} + \eta \mathcal{K}} \left(1 - \eta \eta_\mu\right).
    \label{eq:N_mu_breakeven_transmuter}
\end{equation}
We plot the required number of catalyzed fusion reactions in \Cref{fig:number_of_muonic_reactions} using $\xi_\mrm{circ} = \xi_\mrm{pro} = 0.1$. For electricity-only $\mu$CF systems, we require $N_\mrm{fus,\mu} \gtrsim 415$ for engineering breakeven. With $\ce{^197Au}$ transmutation, we require $N_\mrm{fus,\mu} \gtrsim 195$ for breakeven; with $\ce{^147Pm}$ transmutation we require $N_\mrm{fus,\mu} \gtrsim 70$; with $\ce{^99Mo}$ transmutation we require $N_\mrm{fus,\mu} \gtrsim 0.4$; with $\ce{^225Ac}$ transmutation we require $N_\mrm{fus,\mu} \gtrsim 5 \times 10^{-7}$.

\section{A $\mu$CF System for $\ce{^225Ac}$ Production} \label{sec:ac225}

\iffalse
\begin{table*}[tb!]
\centering
\caption{Comparison of spherical-shell configurations for Ra and RaO blankets under low and high-flux conditions in OpenMC simulations. $\langle p_\mathrm{fus} \rangle$ is the average surface heat flux from fusion wall loading.}
\vspace{4pt}
\begin{tabular}{lccccccc}
\hline
\shortstack{\textbf{Configuration} \\ $\;$}
& \shortstack{$R_\mathrm{in}$ \\ (cm)} 
& \shortstack{$\Delta_R$ \\ (cm)}
& \shortstack{$^{225}$Ac \\ ($\mu$g/yr)}
& \shortstack{$\eta_\mathrm{pro}$ \\ $\;$}
& \shortstack{$\langle p_\mathrm{wall} \rangle^\mu$ \\ (MW/m$^2$)}
& \shortstack{$\langle p_\mathrm{fus} \rangle$ \\ (MW/m$^3$)}
& \shortstack{$\Phi_0$ \\ (n / cm$^2$ s)}
\\
\hline
Low-flux Ra  & 0.10 & 0.65 & 33.5 & 0.0073 & 0.09 & 262 & $1.6\cdot 10^{13}$ \\
Low-flux RaO  & 0.10 & 0.60 & 38.5 & 0.0084 & 0.09 & 262 & $1.6\cdot 10^{13}$ \\
High-flux Ra  & 0.004 & 0.75 & 38.2 & 0.0083 & 11 & $4.1 \cdot 10^6$ & $2\cdot 10^{15}$ \\
High-flux RaO & 0.004 & 0.70 & 44.6 & 0.0097 & 11 & $4.1 \cdot 10^6$ & $2\cdot 10^{15}$ \\
\hline
\end{tabular}
\label{tab:openmcresults}
\end{table*}
\fi

\begin{table*}[tb!]
\centering
\caption{RaO blanket under high-flux conditions in OpenMC simulation. $\langle p_\mathrm{fus} \rangle$ is the average surface heat flux from fusion wall loading.}
\vspace{4pt}
\begin{tabular}{lcccccccc}
\hline
\shortstack{\textbf{Configuration} \\ $\;$}
& \shortstack{$P_\mathrm{fus}$ \\ (Watts)}
& \shortstack{$R_\mathrm{in}$ \\ (cm)} 
& \shortstack{$\Delta_R$ \\ (cm)}
& \shortstack{$^{225}$Ac \\ ($\mu$g/yr)}
& \shortstack{$\eta_\mathrm{pro}$ \\ $\;$}
& \shortstack{$\langle p_\mathrm{wall} \rangle^\mu$ \\ (MW/m$^2$)}
& \shortstack{$\langle p_\mathrm{fus} \rangle$ \\ (MW/m$^3$)}
& \shortstack{$\Phi_0$ \\ (n / cm$^2$ s)}
\\
\hline
High-flux RaO & 564 & 0.089 & 0.62 & 20,480 & 0.0087 & 11 & $1.9\cdot 10^5$ & $2\cdot 10^{15}$ \\
\hline
\end{tabular}
\label{tab:openmcresults}
\end{table*}

%Based on comments from Faris Fakhry, we rederive these arguments without the thin shell approximation. This gives significantly different results at lower muon rates because the thin shell approximation was not valid.

In this section we describe a $\mu$CF neutron source for $\ce{^225Ac}$ \cite{miederer2008realizing,a2011actinium,morgenstern2020supply} production using the transmutation pathway
\begin{equation}
\ce{^{226}Ra(\mrm{n,2n})^{225}Ra} \ce{->[\beta^-][{15\ \text{days}}] ^{225}Ac}.
\end{equation} 
with $\ce{^226Ra}$ feedstock. Given the current scarcity of $\ce{^226Ra}$, we limit ourselves to a total feedstock mass $M_{\mathrm{Ra}} = 10\ \mathrm{g}$, unless mentioned otherwise. A steady-state muon source with power $P_\mu$ and muon production energy cost $E_\mu$ (\Cref{eq:P_mu_eq}) produces
\begin{equation}
\dot N_\mathrm{n}
= N_{\mathrm{fus},\mu} \frac{P_\mu}{E_\mu} f_\mrm{stop},
\label{eq:Nn_main_nothin}
\end{equation}
fusion neutrons per second. We take the $\mu$CF source to be at the center of a spherical $\ce{^{226}Ra}$ shell. To maintain a maximum allowable neutron flux at the inner surface,
\begin{equation}
\Phi_{\max} = 2 \cdot 10^{15}\ \mathrm{n\ cm^{-2}\ s^{-1}},
\end{equation}
the radius of the inner surface $R_{\mathrm{in}}$ is fixed by
\begin{equation}
\dot N_\mathrm{n}
= 4\pi R_{\mathrm{in}}^2 \Phi_{\max},
\qquad
R_{\mathrm{in}}
= \sqrt{\frac{\dot N_\mathrm{n}}{4\pi \Phi_{\max}}}.
\label{eq:Rin_main}
\end{equation}
The $\ce{^{226}Ra}$ blanket occupies a spherical shell of thickness $\Delta R$ between radii $R_{\mathrm{in}}$ and $R_{\mathrm{out}} = R_{\mathrm{in}} + \Delta R$, with mass density $\rho_{\mathrm{Ra}}$. The total feedstock mass is
\begin{equation}
M_{\mathrm{Ra}}
= \rho_{\mathrm{Ra}} V_{\mathrm{shell}}
= \rho_{\mathrm{Ra}} \frac{4\pi}{3}\big(R_{\mathrm{out}}^3 - R_{\mathrm{in}}^3\big),
\label{eq:M_fullgeom}
\end{equation}
where $V_{\mathrm{shell}}$ is the spherical shell volume. Writing $R_{\mathrm{out}} = R_{\mathrm{in}} + \Delta R$ gives
\begin{equation}
M_{\mathrm{Ra}}
= \frac{4\pi \rho_{\mathrm{Ra}}}{3}
\Big[(R_{\mathrm{in}} + \Delta R)^3 - R_{\mathrm{in}}^3\Big],
\label{eq:M_DeltaR_relation}
\end{equation}
which relates the shell thickness $\Delta R$ to the neutron source strength (through $R_{\mathrm{in}}$) and the fixed mass $M_{\mathrm{Ra}}$,
\begin{equation}
\Delta R
= \left[
\left(\frac{\dot{N}_\mathrm{n}}{4\pi\Phi_{\max}}\right)^{3/2}
+ \frac{3M_{\mathrm{Ra}}}{4\pi\rho_{\mathrm{Ra}}}
\right]^{1/3}
-
\left(\frac{\dot{N}_\mathrm{n}}{4\pi\Phi_{\max}}\right)^{1/2}.
\label{eq:DeltaR_final}
\end{equation}
The microscopic (n,2n) cross section of $\ce{^{226}Ra}$ is $\sigma_{(\mathrm{n,2n})}$ and the number density is
\begin{equation}
n_{\mathrm{Ra}} = \frac{\rho_{\mathrm{Ra}} N_A}{A_{\mathrm{Ra}}},
\end{equation}
where $A_{\mathrm{Ra}}$ is the molar mass and $N_A$ is Avogadro’s number. The macroscopic (n,2n) cross section is
\begin{equation}
\Sigma \equiv n_{\mathrm{Ra}} \sigma_{(\mathrm{n,2n})}.
\end{equation}
For radially streaming neutrons traversing the shell, the path length in $\ce{^{226}Ra}$ is $\Delta R$. In a simple exponential attenuation model, the optical depth is
\begin{equation}
\tau = \Sigma \Delta R,
\label{eq:tau_full}
\end{equation}
and the fraction of neutrons that undergo (n,2n) transmutation before exiting the shell is
\begin{equation}
\eta_{\mathrm{pro}}
= 1 - \exp \left( -\tau \right)
= 1 - \exp\big(-\Sigma \Delta R\big).
\label{eq:etapro_full}
\end{equation}
Equations~\eqref{eq:Rin_main} and \eqref{eq:M_DeltaR_relation} determine $R_{\mathrm{in}}$ and $\Delta R$ for a given $\dot N_\mathrm{n}$, $\Phi_{\max}$, $M_{\mathrm{Ra}}$, and $\rho_{\mathrm{Ra}}$. The $\ce{^225Ac}$ production rate is therefore
\begin{equation}
\dot N_{\ce{^225Ac}}
= \eta_{\mathrm{pro}} \dot N_\mathrm{n}
= \big[1 - \exp(-\Sigma \Delta R)\big] \dot N_\mathrm{n},
\label{eq:Pac_full}
\end{equation}
with $\Delta R$ fixed by the spherical geometry relation \eqref{eq:M_DeltaR_relation}. 

\iffalse
\begin{figure}[bt!]
    \centering
    \begin{subfigure}[t]{\textwidth}
    \centering    
    \includegraphics[width=0.99\textwidth]{production_rate_two_techniques.pdf}
    \end{subfigure}
    \caption{OpenMC simulation results for a $P_\mrm{fus} = 1.1$ Watt source with a muon source rate of $2\cdot 10^9$/s.}
    \label{fig:openmc_results}
\end{figure}

\begin{figure*}[tb!]
    \centering
    \begin{subfigure}[t]{\textwidth}
    \centering    
    \includegraphics[width=0.9\textwidth]{two_muCF_RaO_systems.pdf}
    \end{subfigure}
    \caption{(b) Fusion sources and blankets for low (left) and high (right) flux with RaO blankets.}
    \label{fig:openmc_builds}
\end{figure*}
\fi

In \Cref{fig:ac225_production_general}(a), we plot the annual production rate of $\ce{^225Ac}$ in grams versus the muon rate subject to the constraints above. The `high-flux' scenario corresponds to $2\cdot 10^{15}\ \mathrm{n\ cm^{-2}\ s^{-1}}$. The `low-flux' scenario corresponds to $1.6\cdot 10^{12}\ \mathrm{n\ cm^{-2}\ s^{-1}}$, which is the neutron flux a $Q^\mrm{plas} = 0.05$ machine can achieve with $\langle p_\mrm{wall} \rangle = 10$ MW / m$^2$ (see \Cref{fig:pwall_versus_flux}). In \Cref{fig:ac225_production_general}(b) we plot the annual production rate for different feedstock quantities; at lower beam rates annual production scales linearly with $\ce{^226Ra}$ inventory. This is because the blanket is sufficiently thin for most feedstock values that $\ce{^225Ac}$ production does not saturate with increasing feedstock - this is shown by the $\eta_\mathrm{pro}$ values in \Cref{fig:ac225_production_general}(d); even with 10000g of $\ce{^226Ra}$ inventory at most only 10\% of D-T neutrons drive $\ce{^{226}Ra(\mrm{n,2n})^{225}Ra}$ reactions. This means that there are many neutrons available to drive reactions on other feedstock, so an inner $\ce{^226Ra}$ layer may be enclosed within another feedstock layer to produce other medical isotopes \cite{parisi2025production,evitts2025theoretical}. The $\ce{^226Ra}$ blanket layer thickness is shown in \Cref{fig:ac225_production_general}(c).

We now consider a design point that produces roughly twenty milligrams of $\ce{^225Ac}$ per year, approximately 400 times higher than global production in 2024, and validate this choice using an OpenMC \cite{romano2015openmc} simulation. Assuming $N_\mrm{fus,\mu} = 200$, this corresponds to $10^{12}$ muons / second beam rate and $P_\mrm{fus}\approx$564 Watts. We perform an OpenMC simulation with a radium oxide (RaO) blanket, which is used to increase the density of $\ce{^226Ra}$ by approximately 25\%, which increases the $\ce{^225Ac}$ production rate due to a higher number density. As above, the $\ce{^226Ra}$ feedstock inventory is limited to 10 grams. We also surround each blanket in a 20cm graphite reflector, which further increases production by 1-2\% relative to no reflector (results not shown for the no reflector case). The results are summarized in \Cref{tab:openmcresults}. Further work is required to secure the radium supply chain as well as to build systems that efficiently extract $\ce{^225Ac}$ from radium \cite{hagemann1950isolation,sekine1967studies,fitzsimmons2019optimization,Brown2021b}. In this paper we have focused on the radioisotope $\ce{^225Ac}$ - however, it should be noted that there are other radioisotopes with comparable value per neutron with more easily accessible feedstock. These may also be promising candidates for sub-kW class fusion transmuters.

\section{Discussion} \label{sec:discussion}

We have shown that a $\mu$CF system driving transmutation on a blanket feedstock has significantly relaxed conditions for viability, strongly motivating muon source development with higher production rates. From the perspective of valuable radioisotope transmutation, there is far more value in scaling the muon production rate than increasing the number of catalyzed D-T reactions $N_\mrm{fus,\mu}$ by tens of percent or reducing the energy cost per muon $E_\mu$. A unique advantage of $\mu$CF systems is the relatively low (compared with externally heated fusion systems) heat flux corresponding to neutron production - this allows a much higher neutron to heat flux ratio than standard fusion systems.

We showed that with roughly half a kilowatt of muon-catalyzed D-T fusion power, 400 times the global supply (in 2024) of $\ce{^225Ac}$ could be met, representing a significant supply expansion for a radioisotope that faces significant supply shortages. 

While much remains to be done, this work demonstrates that a path towards practical radioisotope production using $\mu$CF systems appears realizable.

\section{Acknowledgements}

We are grateful to F. Fakhry for finding an assumption in the prior version of the manuscript that was not valid in the regime of interest.

\appendix

\section{Muon-catalyzed Fusion Cycle} \label{app:muoncf}

Muons are lost at a rate
\begin{equation}
    \Lambda = \lambda_0 + \omega \phi \lambda_c,
\end{equation}
where $\lambda_0$ is the muon inverse lifetime, $\omega$ is the sticking probability, $\phi$ is the D-T density, and $\lambda_c$ is the effective cycle frequency with dimensions of volume / time. The stopping rate of muons in a D-T target is
\begin{equation}
    R_\mu = I_\mu f_\mrm{stop}
\end{equation}
where $I_\mu$ is the number of muons / seconds. The equation for the number of muons in the target is
\begin{equation}
    \frac{d N_\mu}{d t} = R_\mu - \Lambda N_\mu.
\end{equation}
In muon steady state, we find
\begin{equation}
    N_\mu^\mrm{ss} = \frac{R_\mu}{\Lambda} = \frac{R_\mu}{\lambda_0 + \omega \phi \lambda_c}
\end{equation}
The fusion rate is therefore
\begin{equation}
    \dot{N}_\mrm{n} = N_\mu^\mrm{ss} R_\mrm{fus},
\end{equation}
where
\begin{equation}
    R_\mrm{fus} \equiv \phi \lambda_c,
\end{equation}
is the number of fusion cycles per muon per second. Therefore the fusion rate is
\begin{equation}
    \dot{N}_\mrm{n} = N_\mu^\mrm{ss} \phi \lambda_c = \phi \lambda_c \frac{R_\mu}{\lambda_0 + \omega \phi \lambda_c}.
\end{equation}
We can simplify the fusion rate by defining the average number of fusions per muon,
\begin{equation}
    N_\mrm{fus,\mu} \equiv \frac{R_\mrm{fus}}{\Lambda} = \frac{\phi \lambda_c}{\lambda_0 + \omega \phi \lambda_c}.
    \label{eq:Nfus_mu_appendix}
\end{equation}
This gives the fusion rate
\begin{equation}
    \dot{N}_\mrm{n} = R_\mu N_\mrm{fus,\mu}.
    \label{eq:Ndot_n_appendix}
\end{equation}

\bibliography{Master_EverythingPlasmaBib-2}

\end{document}